\documentclass{osa-article}

\journal{osac}
\usepackage{siunitx}
\usepackage{makecell}
\usepackage{array}
\usepackage{tablefootnote}
\usepackage[utf8x]{inputenc}
\usepackage{xcolor}

\articletype{Research Article}
\begin{document}

\title{A versatile digital approach to laser frequency comb stabilization}

\author{J. K. Shaw,\authormark{1,2} C. Fredrick,\authormark{1,2} and S. A. Diddams\authormark{1,2,*}}

\address{\authormark{1}Department of Physics, University of Colorado, Boulder, Colorado 80309, USA\\
\authormark{2}National Institute of Standards and Technology, 325 Broadway, Boulder, Colorado 80305, USA\\}

\email{\authormark{*}scott.diddams@nist.gov} %% email address is required



\begin{abstract}
We demonstrate the use of a flexible digital servo system for the optical stabilization of both the repetition rate and carrier-envelope offset frequency of a laser frequency comb. The servo system is based entirely on a low-cost field programmable gate array, simple electronic components, and existing open-source software. Utilizing both slow and fast feedback actuators of a commercial mode-locked laser frequency comb, we maintain cycle-slip free locking of optically-derived beatnotes over a 30 hour period, and measure residual phase noise at or below \textasciitilde $0.1$ rad, corresponding to <100 attosecond timing jitter on the optical phase locks. This stability is sufficient for high-precision frequency comb applications, and indicates comparable performance to existing frequency control systems. The modularity of this system allows for it to be easily adapted to suit the servo actuators of a wide variety of laser frequency combs and continuous-wave lasers, reducing cost and complexity barriers, and enabling digital phase control in a wide range of settings.
\end{abstract}

%%%%%%%%%%%%%%%%%%%%%%%%%%  body  %%%%%%%%%%%%%%%%%%%%%%%%%%
\section{\label{sec:level1}Introduction} \label{introduction}

% Comb background paragraph
The first fully-stabilized mode-locked lasers heralded a new age of precision time and frequency metrology that continues to be an active area of research\cite{hall_nobel, hansch_nobel}. Indeed, today's Ti:sapphire, Er:fiber, and Yb-based combs offer repetition rates ranging from 50 \si{\mega \hertz} to 10 \si{\giga \hertz}\cite{diddams_2010_evolvingcomb} and spectra covering 400-2200 \si{\nano \meter}, with high harmonic generation and intra-pulse difference frequency generation demonstrating coherent spectra below 100 \si{\nano \meter} and past 20 \si{\micro \meter} \cite{Gohle_2005_xuvcomb, Cingoz_nat2012_xuv, Kowligy_2019_eos, Timmers_2018optica_dualspect}. On another front, new electro-optic\cite{Kourogi_1994_eomicro, Carlson_2018science_lockedmicrocomb} and microresonator frequency combs
\cite{Kippenberg_2018_dksreview, KippenbergDiddams_combreview, DelHaye2007_combreview}
suggest a future in which frequency comb-based synthesizers and optical clocks will be even more tightly integrated with electronics platforms\cite{Spencer_nature2018_combsynth, newman_combclock}. Together, the wide range of advances will allow small, robust frequency comb platforms to leave the laboratory and find use in optical timekeeping\cite{Predehl_science2012_timetransfer, Lopez2013_mpq_timetransfer, Giorgetta2013_timetransfer, ludlow_revmod2015_time}, spectroscopy\cite{Coddington_opt16_dualcombrev, Picque2019_natpho_combspect}, trace gas sensing\cite{Schliesser_2005_gassensing, Zhang_2013optlet_opocombspect}, optical ranging\cite{Minoshima_2000optappl_ranging, Coddington2009_ranging}, and astronomy\cite{Steinmetz_2008_mpqastro, Wilken_nat2012_astrocomb, Metcalf19_astrospectro}.

% Project motivation
The frequency modes or "teeth" of a frequency comb are described by the familiar comb equation, $\nu_n = n f_{rep} + f_{ceo}$, where the frequency of the $n$th comb tooth $\nu_n$ is determined by the laser's repetition rate $f_{rep}$ and the carrier-envelope-offset frequency $f_{ceo}$ \cite{REICHERT1999_optcomm, Udem2002_metrology}.
Phase-coherent control of $f_{rep}$ and $f_{ceo}$ is essential to the operation of an optical frequency comb and enables its most powerful applications. The cost and complexity of achieving this stability, however, pose a challenge to frequency comb applications. A variety of alternate solutions have been developed to avoid active frequency comb control\cite{Deschenes_oe2010_cwdcs, Ideguchi2014, Truong_oe16_freqref, Zhao_oe2016_freerundcs}, though these approaches exchange the requirements of active feedback for their own complexity in the form of additional lasers and electronics, while still requiring feedback for long-term operation. The rapid proliferation of flexible digital control platforms presents the opportunity to drastically simplify commonly used analog phase control systems, limiting the need for such workarounds, and reducing technical barriers that limit frequency comb applications. 

% Digital servo benefits and examples
In comparison to analog feedback loops, digital systems are less sensitive to electro-magnetic interference\cite{leibrandt_rsi_2015} and have been demonstrated to unwrap and track phase deviations up to $2^{22} \pi$ radians, greatly exceeding the $2\pi$ ambiguity that limits analog systems\cite{dpll_github}. Easily tunable filter parameters and a large and robust capture range are essential for situations that require stabilizing broad linewidth lasers and for frequency comb systems meant to operate in the dynamic environments that exist outside the lab\cite{Leibrandt_opex2011_carcomb, sinclair_compact_comb, Lezius_optica2016_spacecomb, Sinclair_14_opex_outdoorcomb}. Furthermore, digital control can unite the operation and characterization of a frequency comb onto a single platform. With appropriate programming, a field-programmable gate array (FPGA) based locking system can also function as a spectrum analyzer, vector network analyzer, phase noise analyzer, or frequency counter, reducing the overhead cost associated with an optical phase lock.

Digital phase and frequency control of lasers has been employed in research labs for at least two decades\cite{develyntf_1994_olddigitallock, cacciapuoti_rsi2005_phaselock}, and has more recently been demonstrated to achieve robust frequency comb stabilization over extended periods without sacrificing short-term stability\cite{herman_2018pra_opttiming, sinclair_compact_comb, Bluestone_oe17_cwcontrol}. Several laser suppliers now offer digital control systems as standard products for laser phase stabilization\cite{comm_combs}. Still, the cost of these products may put them out of reach of some labs, and they are not always suited to drive the feedback actuators of home-built frequency combs. Flexible, inexpensive platforms for digital stabilization are needed to fully support the diverse applications of laser frequency combs. This will make phase-stabilization feasible in a wider variety of settings, including low-cost education-focused projects.

% Summary
Recently, laser control systems based on the Red Pitaya 125-14 FPGA board\cite{rp_website} have been used to stabilize optical networks\cite{tourignyplante_rsi_2018} and to provide complete signal lock-in modulation and demodulation\cite{luda_2019_rsi}. In this paper, we demonstrate the use of the Red Pitaya and open-source software as a low-cost digital phase locking system for a commercially-available Erbium:fiber-based frequency comb. We report residual phase noise in the locks comparable to values reported by the laser manufacturer, indicating that no significant limitation is imposed by the Red Pitaya. In the long-term, we maintain optical phase locks free of cycle slips for continuous periods exceeding one day, with measurements of frequency stability limited by our radio frequency (RF) standards. We anticipate that this work can serve as a template for servo systems applicable to a wide range of frequency comb sources, making phase stabilization suitable for a growing application space.

\section{Optical Frequency Comb and Control Signal Generation}
\label{beat notes}
To demonstrate the capabilities of the Red Pitaya FPGA-based frequency control, we employ a commercial Er:fiber mode-locked laser (Menlo Systems FC1500-250-ULN) with a 250 \si{\mega \hertz} repetition rate\cite{hansel_2017_menlo}. All optical locking signals are derived from a polarization-maintaining (PM) fiber output of 30 mW average power and 40 \si{\nano \meter} spectral bandwidth centered at 1560 \si{\nano \meter}. In the following, we describe the generation of $f_{ceo}$ via f-2f interferometry, as well as the control of $f_{rep}$ through the simultaneous optical stabilization of a single mode, $\nu_N$, of the frequency comb. We have also used the same FPGA hardware to stabilize $f_{rep}$ directly in the microwave domain.

\subsection{Optical Beat Signals}
\begin{figure}
\centering
\includegraphics[scale=0.30]{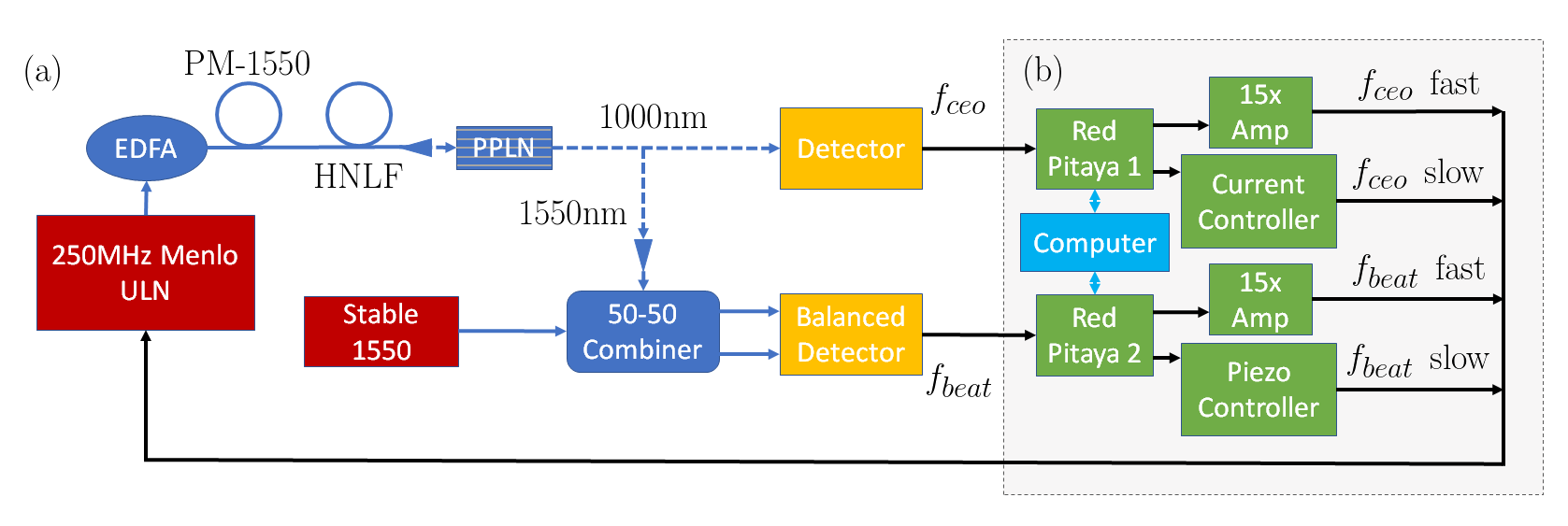} %0.35
\caption{\label{fig:f_2f} (a) Block diagram showing the generation and detection of $f_{ceo}$ and $f_{beat}$. Femtosecond pulses from the laser are amplified in Erbium-doped fiber and compressed in PM-1550 fiber to produce an octave-spanning spectrum in HNLF. A free-space, dual-ridge, PPLN waveguide with poling periods optimized for second harmonic generation at 1005 \si{\nano \meter} doubles light at 2 \si{\micro \meter} to 1 \si{\micro \meter}. We detect $f_{ceo}$ at 1 \si{\micro \meter}, while 1550 \si{\nano \meter} supercontinuum light is mixed with light from a 1550 \si{\nano \meter} laser cavity to produce $f_{beat}$. EDFA: erbium-doped fiber amplifier, HNLF: highly non-linear fiber, PPLN: periodically-poled lithium niobate. (b) Block diagram showing the electronics feedback chain used to lock $f_{ceo}$ and $f_{beat}$.}
\end{figure}

Conventional techniques are employed to generate the error signals (optical beat notes) that are subsequently processed with the FPGA to control the frequency comb. Generation and measurement of $f_{ceo}$ and an optical beat note $f_{beat}$ between a continuous wave (CW) and a single comb tooth\cite{REICHERT1999_optcomm, diddams_2000_link, Diddams_2001_hgclock} are critical steps in frequency comb stabilization that have been well-described in the literature.  With reference to Fig. \ref{fig:f_2f}(a), here we briefly review the details relevant to our experiments.

\begin{figure}
\centering
\includegraphics[scale=0.35]{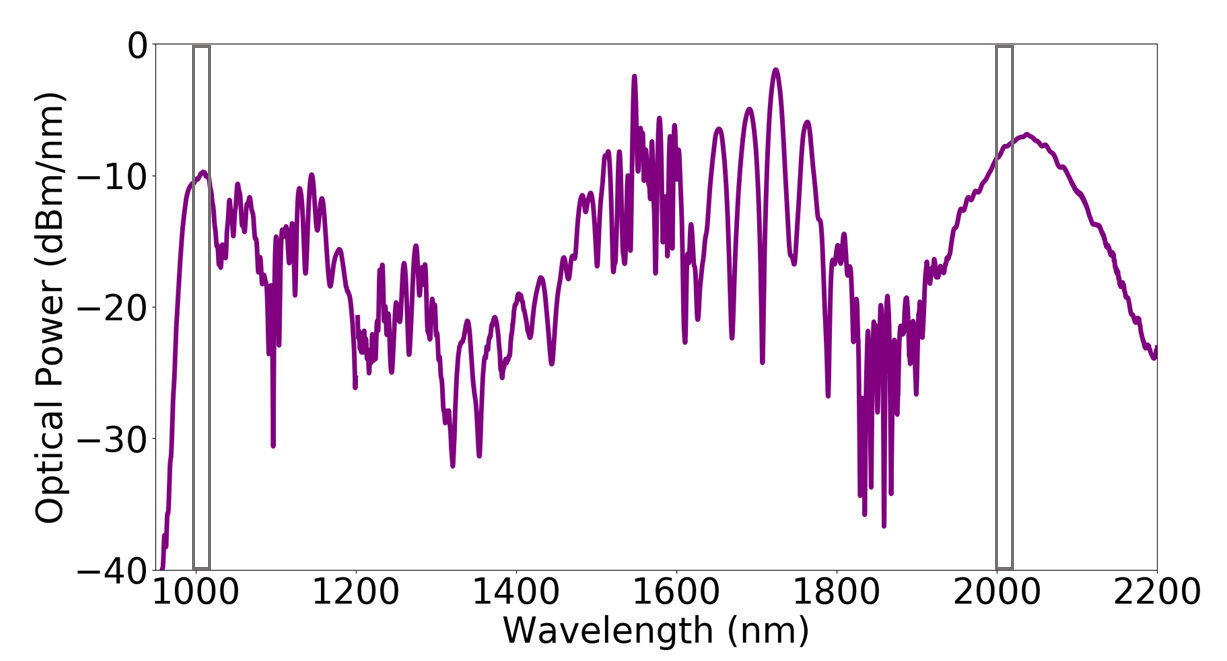}
\caption{\label{fig:supercontinuum} Supercontinuum produced in HNLF. Vertical boxes highlight portions of the octave-spanning spectrum used to produce the $f_{ceo}$ beatnote. To broaden sufficiently, we use 80 fs input pulses with \textasciitilde1 nJ pulse energy and 56 cm of PM-HNLF with dispersion 5.7 ps/(nm km) and non-linearity 10.5 (W km)$^{-1}$.}
\end{figure}

\begin{figure}
\centering
\includegraphics[scale=0.33]{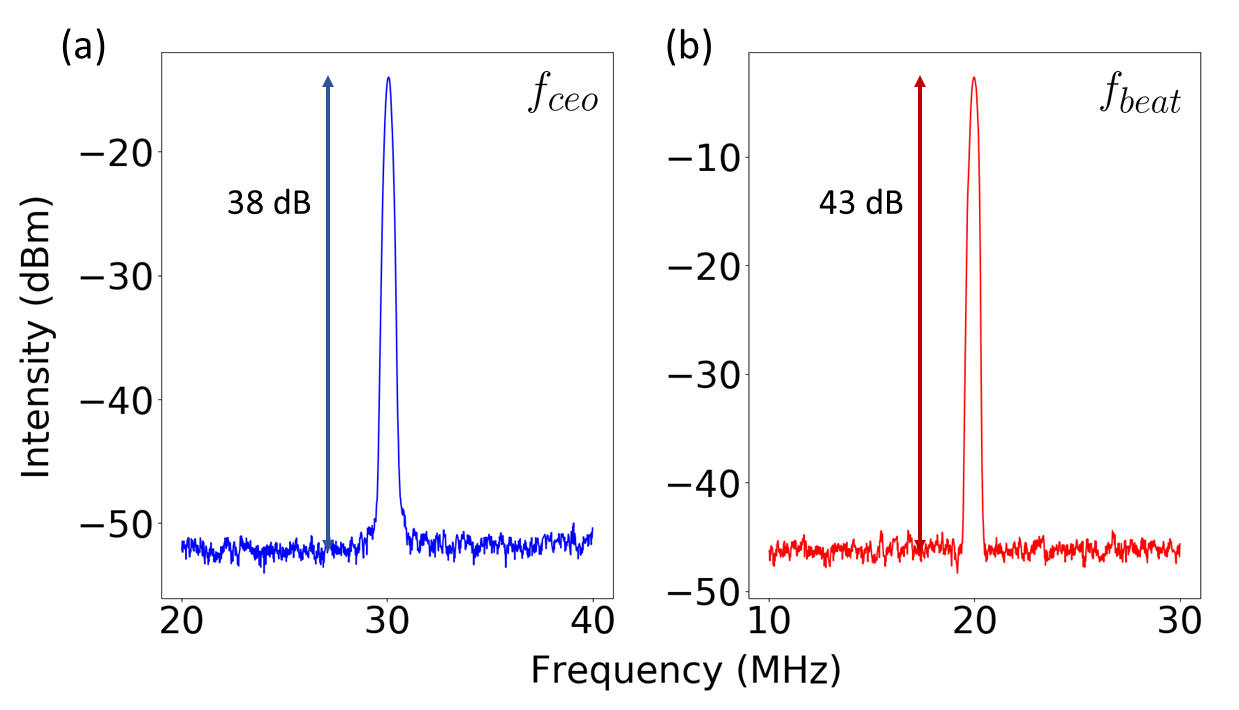}
\caption{\label{fig:unlocked_beatnotes_1} Free-running beat notes measured at 300 kHz resolution bandwidth. a. The $f_{ceo}$ heterodyne shows 38 dB SNR. b. The $f_{beat}$ heterodyne shows 43 dB SNR.}
\end{figure}

To stabilize $f_{ceo}$, we amplify the pulses output from the Er:fiber laser and spectrally-broaden them to octave bandwidth using polarization-maintaining highly nonlinear optical fiber (PM-HNLF)\cite{washburn_hnlfoctave}. The resulting spectrum is shown in Fig.~\ref{fig:supercontinuum}. The output of the PM-HNLF is coupled into a periodically-poled lithium niobate (PPLN) ridge-waveguide\cite{Nishida_2003elett_nttppln} with poling period optimized for second harmonic generation at 1005 \si{\nano \meter}. Supercontinuum comb light at 1 \si{\micro \meter} as well as 2 \si{\micro \meter} comb light that is frequency doubled in the waveguide are bandpass filtered before impinging on an InGaAs photodetector. The resulting $f_{ceo}$ beatnote is shown in Fig.~\ref{fig:unlocked_beatnotes_1}(a), with 38 dB signal-to-noise ratio (SNR) at 300 kHz resolution bandwidth (RBW).

To stabilize $f_{rep}$, we heterodyne a single comb mode near 1550 \si{\nano \meter} with a cavity-stabilized 1550 \si{\nano \meter} CW laser to produce the RF signal $f_{beat}$ with 43 dB SNR (300 kHz RBW), shown in Fig. \ref{fig:unlocked_beatnotes_1}(b). A fiber-coupled band-pass filter is used to reduce the comb bandwidth on the detector to 1 \si{\nano \meter} before combining with the 1550 \si{\nano \meter} CW light in a four-port fiber-coupled beam splitter.  The two output ports of the combiner are directed to a balanced photodiode, which outputs $f_{beat}$. We note that stabilizing $f_{beat}$ instead of directly stabilizing $f_{rep}$ provides increased leverage to control the comb's repetition rate and achieve optical coherence.

\subsection{FPGA Hardware and Electronic Feedback Control}
We use two Red Pitaya 125-14 FPGA boards to sample and process $f_{ceo}$ and $f_{beat}$, providing feedback to stabilize the frequency comb. The Red Pitayas are externally clocked by a 125 \si{\mega \hertz} maser-referenced signal distributed by a low-noise clock buffer (LTC6954). Each board has two analog inputs and outputs with 14-bit resolution analog to digital (ADC) and digital to analog (DAC) converters. We make a single hardware modification to the Red Pitaya boards to enable external clocking.

The architecture of the FPGA and python code used to control and monitor the servo loops are discussed in detail by Tourigny-Plante et al.\cite{tourignyplante_rsi_2018}. In brief, the FPGA mixes the digitized analog input signal with digitally synthesized I/Q reference frequencies to detect the phase error. A term proportional to the frequency error is then calculated by taking the difference of the phase error from one clock cycle to the next. This frequency error is passed into the loop filter which contains proportional, integral, double-integral and differential terms ($\rm{PII^2D}$). The output of the loop filter is sent to the DAC to produce an analog signal that drives the laser actuators. In addition to running PID loops, the board records the frequency error and the voltage output of both DACs and displays the lock stability and output range on the python-based graphic user interface (GUI). Using the GUI as a display, the Red Pitaya can also act as a spectrum analyzer, vector network analyzer, phase noise analyzer, and frequency counter. All firmware and software employed in this work are available for download from a fork of the Github repository linked in Ref. \citen{dpll_github}.

Each FPGA board stabilizes either $f_{ceo}$ or $f_{beat}$ through control of both a fast and slow feedback source on the laser. This electronics chain is shown in Fig. \ref{fig:f_2f}(b).
The Er:fiber mode-locked laser has two EOMs to provide fast feedback on $f_{ceo}$\cite{menlo_cleo2017_ceo} and $f_{beat}$. We amplify the -1 to 1 V standard output of the Red Pitaya DAC with a 15x non-inverting amplifier (AD744) to 30 V of tuning range on the EOMs. The characteristics of the full electronics chain used to provide feedback are described in Table \ref{tab:feedback}.

In addition to providing feedback to the EOMs, the output of the fast PID loop serves as the error signal for the slow servo locks. These auxiliary servos have a greater range, but lower bandwidth and are used to keep the EOM servos at the center of their range. By setting a large time constant on the integrator term of these loops, we create slow, secondary feedback that allows for long-term continuous locking without interfering with the performance of the fast locks. In order to reach these time constants, the bit depth of the integral terms' gain was doubled with respect to the configuration in Tourigny-Plante et al.\cite{tourignyplante_rsi_2018}. In the case of the Red Pitaya that stabilizes $f_{ceo}$, the slow feedback (\textasciitilde 10 Hz) drives a current controller (Thorlabs LDC200CV) to adjust the temperature of the EOM that stabilizes $f_{ceo}$. For slow stabilization of $f_{beat}$, the Red Pitaya's output serves as modulation for a 0-150 V piezo driver (Thorlabs KPZ101) controlling a PZT within the laser cavity. While the EOM temperature tunes $f_{ceo}$ nearly independently of $f_{rep}$, the PZT voltage changes $f_{ceo}$ as well as the intended repetition rate. This effect is much greater than the natural drift of $f_{ceo}$, meaning that the slow feedback on $f_{ceo}$ serves largely to compensate for the secondary effects of the PZT. For both slow servos, the standard output voltage range of the Red Pitaya's DACs is sufficient to keep the fast EOM servos within their range for several days. 

\begin{table}
\centering
\caption{\label{tab:feedback}Feedback Chain Characteristics for Optical Locks.}
\begin{tabular}{|>{\centering\arraybackslash}m{2.25cm}|>{\centering\arraybackslash}m{2cm}>{\centering\arraybackslash}m{2cm}|>{\centering\arraybackslash}m{2cm}>{\centering\arraybackslash}m{2cm}|}
\hline
Control Signal&\multicolumn{2}{c|}{$f_{ceo}$}&\multicolumn{2}{c|}{$f_{beat}$}\\ \hline
Speed&fast&slow&fast&slow \\
Bandwidth & \textasciitilde100 kHz & 10 Hz & \textasciitilde100 kHz & 1 Hz \\ 
Modulation Control&FPGA DAC \hspace{0.5cm} -1 to 1 V&FPGA DAC \hspace{0.5cm} -1 to 1 V&FPGA DAC \hspace{0.5cm} -1 to 1 V&FPGA DAC \hspace{0.5cm} 0 to 1 V \\
Modulation Source & 15x \hspace{0.5cm} Amplifier & Current Controller\tablefootnote{Thorlabs LDC200CV Current Controller, 0 - 20 mA total range.} & 15x \hspace{0.5cm} Amplifier & HV Amplifier\tablefootnote{Thorlabs KPZ101 Piezo Controller, 0 - 100 V total range.} \\
Modulation Range & $\pm 15$ V & 4 mA & $\pm 15$ V & 15 V \\
Actuator & EOM & TEC & EOM & PZT \\
Actuator Limits & $\pm 48$ V & $\pm 30$ mA & $\pm 48$ V & -15 - 150 V \\
Tuning Range & 11.4 \si{\mega \hertz} & 66 \si{\mega \hertz} & 3.8 \si{\mega \hertz} & 36 \si{\mega \hertz} \\
\hline
\end{tabular}
\end{table}

% What extent of this is going to go in the introduction? I am going to assume general RP stuff is covered in the intro.
\section{Results and Characterization of Frequency Comb Stabilization } \label{stabilization}
\begin{figure}
\centering
\includegraphics[scale=0.33]{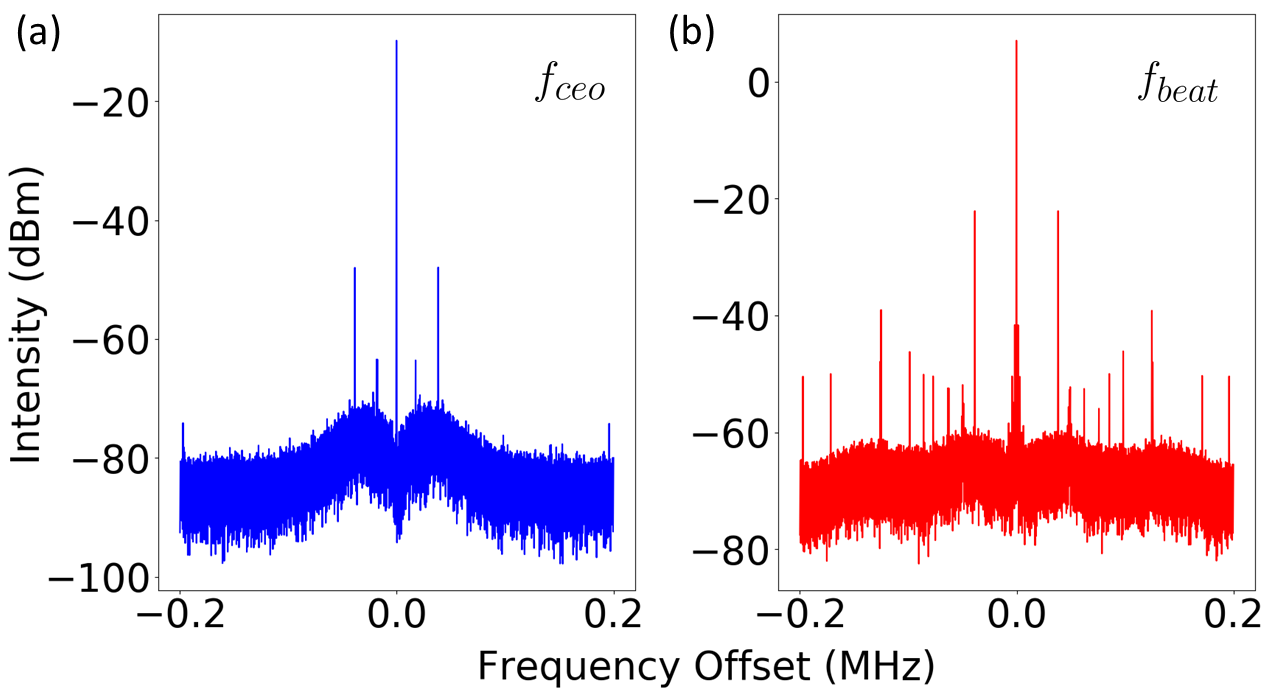}
\caption{\label{fig:locked_beatnotes_1} a. Stabilized $f_{ceo}$ beatnote at 10 Hz resolution bandwidth. b. Stabilized $f_{beat}$ beatnote at 10 Hz resolution bandwidth.}
\end{figure}

When the FPGA (Red Pitaya) servo loops are closed, the linewidths of both $f_{ceo}$ and $f_{beat}$ are reduced to narrow coherent carriers.  Figure~\ref{fig:locked_beatnotes_1} shows the phase-locked $f_{ceo}$ and $f_{beat}$ beatnotes at 10 Hz resolution bandwidth. Further analysis of these beats provides characterization of both the short- and long-term stability of the phase locks.

As a consequence of unifying feedback and diagnostics on a single platform, all measurements made by the FPGA board are completely in-loop with the phase locks. As a result, we observe artificially low noise at low frequencies where the loop gain is the highest, recording residual frequency errors in the 100s of \si{\nano \hertz} on the internal frequency counter. Measurements made at higher frequencies, such as phase noise, are less affected. To avoid this bias entirely, we report phase noise and average frequency measurements recorded by instruments out of loop to the phase locks, but still referenced to a shared maser clock source.

\subsection{Short-term locking behavior}
\begin{figure}
\centering
\includegraphics[scale=0.45]{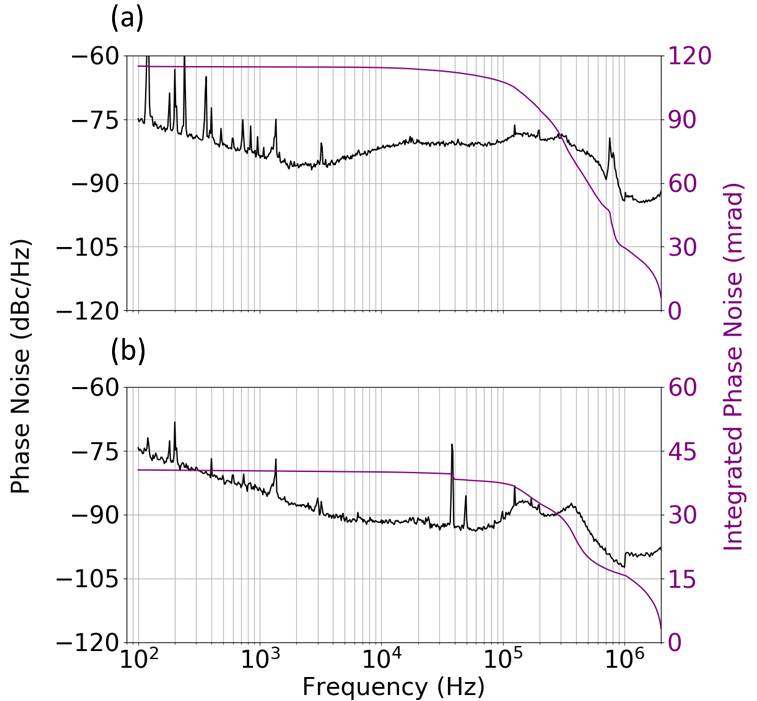}
\caption{\label{fig:ipn_1} a. Phase noise (black) and integrated phase noise (purple) on $f_{beat}$. The total integrated phase noise from 100 Hz to 2 \si{\mega \hertz} is 114 mrad. b. Phase noise (black) and integrated phase noise (purple) on $f_{ceo}$. The total integrated phase noise from 100 Hz to 2 \si{\mega \hertz} is 40 mrad.}
\end{figure}

Figure~\ref{fig:ipn_1} shows the phase noise of $f_{ceo}$ and $f_{beat}$ measured from 100 Hz to 2 \si{\mega \hertz} with an Agilent MXA N9020A signal analyzer. We measured the integrated phase noise over this bandwidth to be 114 mrad for $f_{ceo}$ and 41 mrad for $f_{beat}$. The laser's manufacturer reports values of 85 mrad and 43 mrad, indicating that the Red Pitaya FPGA servo system performs comparably to feedback systems designed specifically for the laser. We calculate the corresponding integrated timing jitter to be 70 attoseconds, within the requirements of broadband precision dual-comb spectroscopy and well below the present requirements of optical ranging and fiber-network applications. Both signals have an effective feedback bandwidth of \textasciitilde100 kHz, limited by 565 \si{\nano \second} latency of the FPGA platform\cite{tourignyplante_rsi_2018} and delays within the RF filtering and amplification chain.

\subsection{Long-term locking behavior}
\begin{figure}
\centering
\includegraphics[scale=0.40]{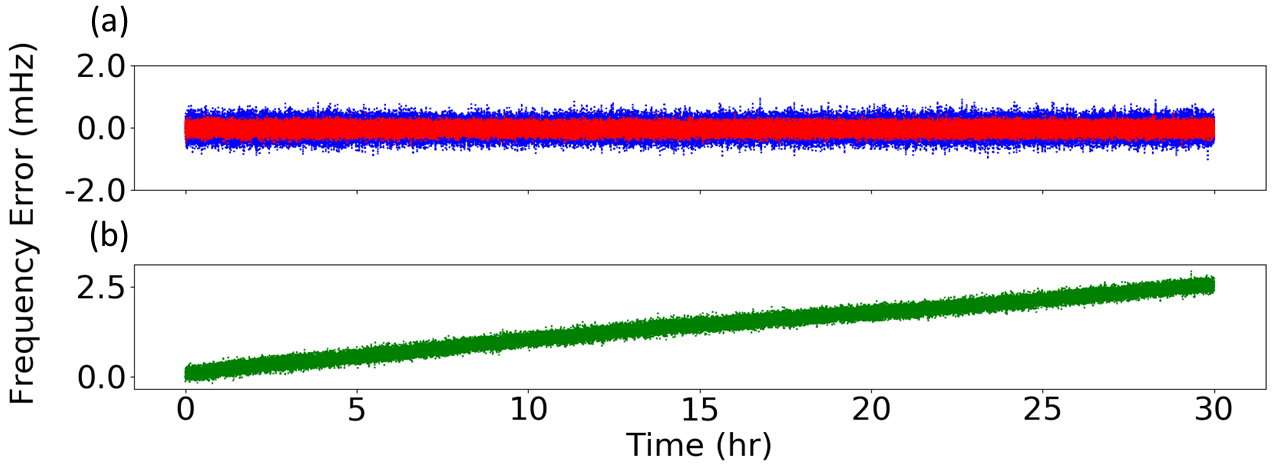} %0.51
\caption{\label{fig:freq_offsets_1} Frequency error from set-points over 30 hour period. (a) Frequency error in optical beat notes $f_{ceo}$ (blue) and $f_{beat}$ (red). The standard deviations of  $f_{ceo}$ and $f_{beat}$ are 0.2 \si{\milli \hertz} and 0.1 \si{\milli \hertz}, respectively. (b) Frequency error in $f_{rep}$. After correcting for a linear drift, the standard deviation of $f_{rep}$ is calculated as 0.1 \si{\milli \hertz}. All measurements were made by external frequency counters.}
\end{figure}

To demonstrate the consistency of the digital stabilization platform, we leave the comb locked for extended intervals. Figure \ref{fig:freq_offsets_1} shows the frequency deviation of $f_{ceo}$, $f_{beat}$, and $f_{rep}$ measured with Agilent 53132A frequency counters over a 30 hour period. These data demonstrate that the servos can be free of cycle slips over extended timescales.

The change in $f_{rep}$ in Fig. \ref{fig:freq_offsets_1}(b) shows the slow drift of the 1550 \si{\nano \meter} optical cavity against the long-term stability of the maser signal. The 2.5 \si{\milli \hertz} change in the repetition rate over 30 hours corresponds to a drift of the 1550 \si{\nano \meter} optical frequency of only 20 \si{\milli \hertz} per second. We calculate the frequency uncertainty at 1 s to be 0.2 \si{\milli \hertz} and 0.1 \si{\milli \hertz} for $f_{ceo}$ and $f_{beat}$, respectively. The uncertainty of these measurements is limited at this level by both the reported counter uncertainty, and the instability of the maser-referenced synthesizer used to clock the Red Pitayas.

\begin{figure}
\centering
\includegraphics[scale=0.25]{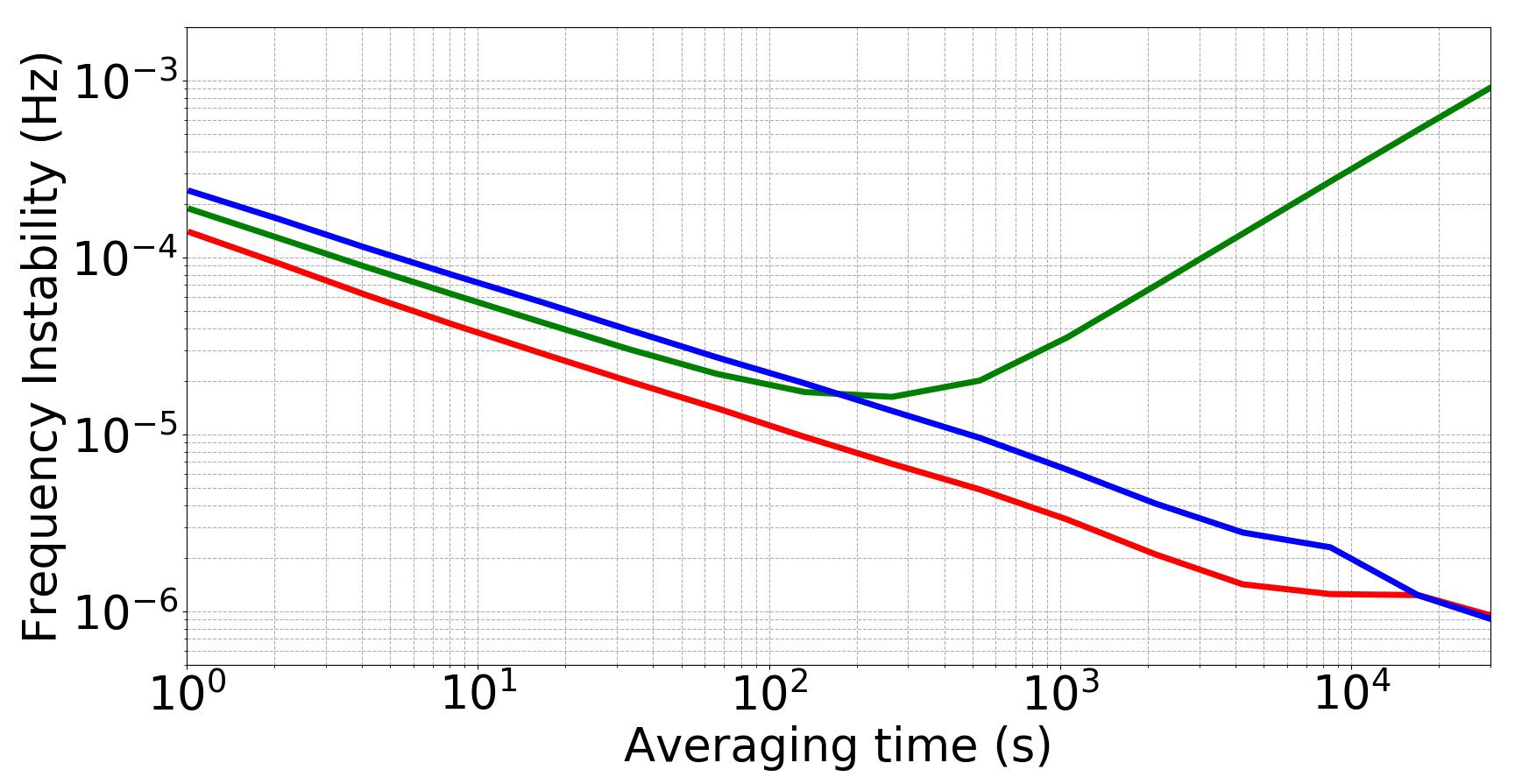}
\caption{\label{fig:all_adevs} Allan Deviations for $f_{ceo}$ (blue), $f_{beat}$ (red), and $f_{rep}$ (green), during an optical lock. For small averaging times, all three frequencies are limited by the RF source used to clock the phase-locks and counters. For averaging times exceeding \textasciitilde100 s, the slow drift of the 1550 \si{\nano \meter} cavity limits the stability of $f_{rep}$.}
\end{figure}

Figure \ref{fig:all_adevs} shows the calculated Allan Deviations of $f_{rep}$, $f_{ceo}$, and $f_{beat}$ for the same 30 hour period.
For small averaging times, the measurement of all three frequencies is limited by the RF electronics as previously mentioned. The stability of $f_{rep}$ provides an absolute measurement of the 1550 \si{\nano \meter} cavity against the hydrogen maser, showing that the drift of the repetition rate exceeds the uncertainty of the counting electronics for averaging times greater than \textasciitilde100s. Optically derived $f_{ceo}$ (blue) and $f_{beat}$ (red) are in-loop measurements effectively indicating the quality of the servo, and continue to fall off like $1/\sqrt{\tau}$.

\subsection{Direct Phase Locking of $f_{rep}$}
Our locking set-up is easily reconfigured to stabilize the repetition rate directly in the RF domain. While locking the repetition rate at a low harmonic adds noise in the optical domain, it does not require a stable optical source, and demonstrates the flexibility of the digital locking platform.

When implementing an RF lock, we mix the second harmonic of the 250 \si{\mega \hertz} repetition rate (obtained from a photodiode internal to the laser) with a synthesized 480 \si{\mega \hertz} signal to produce a 20 \si{\mega \hertz} beat note. Just as with the optical lock, this signal can be fed into the Red Pitaya and locked with appropriate settings on the PID loop. Switching between optical and RF locks of the repetition rate requires us only to change the Red Pitaya's ADC input and import saved PID settings to the Red Pitaya through the GUI, a fast and reproducible process. The sensitivity of the RF and optical beat notes to the $f_{rep}$ EOM's voltage differs by a factor of \textasciitilde400,000, in good agreement with the values predicted by the comb equation, and demonstrating the increased sensitivity and multiplicative advantage of the optical lock.

\section{Conclusions} \label{conclusion}
We presented a digital servo system that can be easily configured to stabilize a commercial frequency comb with EOM, PZT, and current feedback sources. The measured phase noise values of 41 mrad and 114 mrad on $f_{beat}$ and $f_{ceo}$, respectively, are in good agreement with the values reported by the manufacturer. The corresponding timing jitter of 70 attoseconds is suitable for the most demanding applications of combs in timing and broadband spectroscopy. Cycle-slip free locking is maintained for 30 hours. The low cost and simplicity of this approach make it appropriate for student-driven projects. We hope that this servo design can be easily adapted to a variety of mode-locked lasers, reducing the technical and financial barriers to optically stabilizing combs.

\section{Acknowledgements}
The authors are very grateful to J.-D. Deschenes for his valuable contributions to this project.  In addition, we thank I. Coddington and F. Quinlan for their comments on this manuscript. We would also like to thank D. Herman, A. Kowligy, D. Lesko, A. Lind, N. Newbury, and H. Timmers for fruitful technical discussion.
This work is supported by NIST and NSF AST 1310875. The use of specific trade names in this manuscript  does not constitute an endorsement by NIST. NIST is an agency of the US Government and this work is not subject to copyright in the US.

%%%%%%%%%%%%%%%%%%%%%%% References %%%%%%%%%%%%%%%%%%%%%%%%%

%%%%%%%%%% If using BibTeX:
\bibliography{main_bib}

\providecommand{\noopsort}[1]{}\providecommand{\singleletter}[1]{#1}%
\begin{thebibliography}{10}
\newcommand{\enquote}[1]{``#1''}

\bibitem{hall_nobel}
J.~L. Hall, \enquote{Nobel lecture: Defining and measuring optical
  frequencies,} {\protect\JournalTitle{Rev. Mod. Phys.}} \textbf{78},
  1279--1295 (2006).

\bibitem{hansch_nobel}
T.~W. H\"ansch, \enquote{Nobel lecture: Passion for precision,}
  {\protect\JournalTitle{Rev. Mod. Phys.}} \textbf{78}, 1297--1309 (2006).

\bibitem{diddams_2010_evolvingcomb}
S.~A. Diddams, \enquote{The evolving optical frequency comb,}
  {\protect\JournalTitle{J. Opt. Soc. Am. B}} \textbf{27}, B51--B62 (2010).

\bibitem{Gohle_2005_xuvcomb}
C.~Gohle, T.~Udem, M.~Herrmann, J.~Rauschenberger, R.~Holzwarth, H.~A.
  Schuessler, F.~Krausz, and T.~W. H{\"a}nsch, \enquote{A frequency comb in the
  extreme ultraviolet,} {\protect\JournalTitle{Nature}} \textbf{436}, 234--237
  (2005).

\bibitem{Cingoz_nat2012_xuv}
A.~Cing{\"o}z, D.~C. Yost, T.~K. Allison, A.~Ruehl, M.~E. Fermann, I.~Hartl,
  and J.~Ye, \enquote{Direct frequency comb spectroscopy in the extreme
  ultraviolet,} {\protect\JournalTitle{Nature}} \textbf{482}, 68 EP -- (2012).

\bibitem{Kowligy_2019_eos}
A.~S. Kowligy, H.~Timmers, A.~J. Lind, U.~Elu, F.~C. Cruz, P.~G. Schunemann,
  J.~Biegert, and S.~A. Diddams, \enquote{Infrared electric field sampled
  frequency comb spectroscopy,} {\protect\JournalTitle{Science Advances}}
  \textbf{5} (2019).

\bibitem{Timmers_2018optica_dualspect}
H.~Timmers, A.~Kowligy, A.~Lind, F.~C. Cruz, N.~Nader, M.~Silfies, G.~Ycas,
  T.~K. Allison, P.~G. Schunemann, S.~B. Papp, and S.~A. Diddams,
  \enquote{Molecular fingerprinting with bright, broadband infrared frequency
  combs,} {\protect\JournalTitle{Optica}} \textbf{5}, 727--732 (2018).

\bibitem{Kourogi_1994_eomicro}
M.~Kourogi, T.~Enami, and M.~Ohtsu, \enquote{A monolithic optical frequency
  comb generator,} {\protect\JournalTitle{Photonics Technology Letters, IEEE}}
  \textbf{6}, 214 -- 217 (1994).

\bibitem{Carlson_2018science_lockedmicrocomb}
D.~R. Carlson, D.~D. Hickstein, W.~Zhang, A.~J. Metcalf, F.~Quinlan, S.~A.
  Diddams, and S.~B. Papp, \enquote{Ultrafast electro-optic light with subcycle
  control,} {\protect\JournalTitle{Science}} \textbf{361}, 1358--1363 (2018).

\bibitem{Kippenberg_2018_dksreview}
T.~J. Kippenberg, A.~L. Gaeta, M.~Lipson, and M.~L. Gorodetsky,
  \enquote{Dissipative kerr solitons in optical microresonators,}
  {\protect\JournalTitle{Science}} \textbf{361} (2018).

\bibitem{KippenbergDiddams_combreview}
T.~J. Kippenberg, R.~Holzwarth, and S.~A. Diddams,
  \enquote{Microresonator-based optical frequency combs,}
  {\protect\JournalTitle{Science}} \textbf{332}, 555--559 (2011).

\bibitem{DelHaye2007_combreview}
P.~Del'Haye, A.~Schliesser, O.~Arcizet, T.~Wilken, R.~Holzwarth, and T.~J.
  Kippenberg, \enquote{Optical frequency comb generation from a monolithic
  microresonator,} {\protect\JournalTitle{Nature}} \textbf{450}, 1214 EP --
  (2007).

\bibitem{Spencer_nature2018_combsynth}
D.~T. Spencer, T.~Drake, T.~C. Briles, J.~Stone, L.~C. Sinclair, C.~Fredrick,
  Q.~Li, D.~Westly, B.~R. Ilic, A.~Bluestone, N.~Volet, T.~Komljenovic,
  L.~Chang, S.~H. Lee, D.~Y. Oh, M.-G. Suh, K.~Y. Yang, M.~H.~P. Pfeiffer,
  T.~J. Kippenberg, E.~Norberg, L.~Theogarajan, K.~Vahala, N.~R. Newbury,
  K.~Srinivasan, J.~E. Bowers, S.~A. Diddams, and S.~B. Papp, \enquote{An
  optical-frequency synthesizer using integrated photonics,}
  {\protect\JournalTitle{Nature}} \textbf{557}, 81--85 (2018).

\bibitem{newman_combclock}
Z.~L. Newman, V.~Maurice, T.~Drake, J.~R. Stone, T.~C. Briles, D.~T. Spencer,
  C.~Fredrick, Q.~Li, D.~Westly, B.~R. Ilic, B.~Shen, M.-G. Suh, K.~Y. Yang,
  C.~Johnson, D.~M.~S. Johnson, L.~Hollberg, K.~J. Vahala, K.~Srinivasan, S.~A.
  Diddams, J.~Kitching, S.~B. Papp, and M.~T. Hummon, \enquote{Architecture for
  the photonic integration of an optical atomic clock,}
  {\protect\JournalTitle{Optica}} \textbf{6}, 680--685 (2019).

\bibitem{Predehl_science2012_timetransfer}
K.~Predehl, G.~Grosche, S.~M.~F. Raupach, S.~Droste, O.~Terra, J.~Alnis,
  T.~Legero, T.~W. H{\"a}nsch, T.~Udem, R.~Holzwarth, and H.~Schnatz,
  \enquote{A 920-kilometer optical fiber link for frequency metrology at the
  19th decimal place,} {\protect\JournalTitle{Science}} \textbf{336}, 441--444
  (2012).

\bibitem{Lopez2013_mpq_timetransfer}
O.~Lopez, A.~Kanj, P.-E. Pottie, D.~Rovera, J.~Achkar, C.~Chardonnet,
  A.~Amy-Klein, and G.~Santarelli, \enquote{Simultaneous remote transfer of
  accurate timing and optical frequency over a public fiber network,}
  {\protect\JournalTitle{Applied Physics B}} \textbf{110}, 3--6 (2013).

\bibitem{Giorgetta2013_timetransfer}
F.~R. Giorgetta, W.~C. Swann, L.~C. Sinclair, E.~Baumann, I.~Coddington, and
  N.~R. Newbury, \enquote{Optical two-way time and frequency transfer over free
  space,} {\protect\JournalTitle{Nature Photonics}} \textbf{7}, 434 EP --
  (2013).

\bibitem{ludlow_revmod2015_time}
A.~D. Ludlow, M.~M. Boyd, J.~Ye, E.~Peik, and P.~O. Schmidt, \enquote{Optical
  atomic clocks,} {\protect\JournalTitle{Rev. Mod. Phys.}} \textbf{87},
  637--701 (2015).

\bibitem{Coddington_opt16_dualcombrev}
I.~Coddington, N.~Newbury, and W.~Swann, \enquote{Dual-comb spectroscopy,}
  {\protect\JournalTitle{Optica}} \textbf{3}, 414--426 (2016).

\bibitem{Picque2019_natpho_combspect}
N.~Picqu{\'e} and T.~W. H{\"a}nsch, \enquote{Frequency comb spectroscopy,}
  {\protect\JournalTitle{Nature Photonics}} \textbf{13}, 146--157 (2019).

\bibitem{Schliesser_2005_gassensing}
A.~Schliesser, M.~Brehm, F.~Keilmann, and D.~W. van~der Weide,
  \enquote{Frequency-comb infrared spectrometer for rapid, remote chemical
  sensing,} {\protect\JournalTitle{Opt. Express}} \textbf{13}, 9029--9038
  (2005).

\bibitem{Zhang_2013optlet_opocombspect}
Z.~Zhang, T.~Gardiner, and D.~T. Reid, \enquote{Mid-infrared dual-comb
  spectroscopy with an optical parametric oscillator,}
  {\protect\JournalTitle{Opt. Lett.}} \textbf{38}, 3148--3150 (2013).

\bibitem{Minoshima_2000optappl_ranging}
K.~Minoshima and H.~Matsumoto, \enquote{High-accuracy measurement of 240-m
  distance in an optical tunnel by use of a compact femtosecond laser,}
  {\protect\JournalTitle{Appl. Opt.}} \textbf{39}, 5512--5517 (2000).

\bibitem{Coddington2009_ranging}
I.~Coddington, W.~C. Swann, L.~Nenadovic, and N.~R. Newbury, \enquote{Rapid and
  precise absolute distance measurements at long range,}
  {\protect\JournalTitle{Nature Photonics}} \textbf{3}, 351 EP -- (2009).
  Article.

\bibitem{Steinmetz_2008_mpqastro}
T.~Steinmetz, T.~Wilken, C.~Araujo-Hauck, R.~Holzwarth, T.~W. H{\"a}nsch,
  L.~Pasquini, A.~Manescau, S.~D{\textquoteright}Odorico, M.~T. Murphy,
  T.~Kentischer, W.~Schmidt, and T.~Udem, \enquote{Laser frequency combs for
  astronomical observations,} {\protect\JournalTitle{Science}} \textbf{321},
  1335--1337 (2008).

\bibitem{Wilken_nat2012_astrocomb}
T.~Wilken, G.~L. Curto, R.~A. Probst, T.~Steinmetz, A.~Manescau, L.~Pasquini,
  J.~I. Gonz{\'a}lez~Hern{\'a}ndez, R.~Rebolo, T.~W. H{\"a}nsch, T.~Udem, and
  R.~Holzwarth, \enquote{A spectrograph for exoplanet observations calibrated
  at the centimetre-per-second level,} {\protect\JournalTitle{Nature}}
  \textbf{485}, 611 EP -- (2012).

\bibitem{Metcalf19_astrospectro}
A.~J. Metcalf, T.~Anderson, C.~F. Bender, S.~Blakeslee, W.~Brand, D.~R.
  Carlson, W.~D. Cochran, S.~A. Diddams, M.~Endl, C.~Fredrick, S.~Halverson,
  D.~D. Hickstein, F.~Hearty, J.~Jennings, S.~Kanodia, K.~F. Kaplan, E.~Levi,
  E.~Lubar, S.~Mahadevan, A.~Monson, J.~P. Ninan, C.~Nitroy, S.~Osterman, S.~B.
  Papp, F.~Quinlan, L.~Ramsey, P.~Robertson, A.~Roy, C.~Schwab, S.~Sigurdsson,
  K.~Srinivasan, G.~Stefansson, D.~A. Sterner, R.~Terrien, A.~Wolszczan, J.~T.
  Wright, and G.~Ycas, \enquote{Stellar spectroscopy in the near-infrared with
  a laser frequency comb,} {\protect\JournalTitle{Optica}} \textbf{6}, 233--239
  (2019).

\bibitem{REICHERT1999_optcomm}
J.~Reichert, R.~Holzwarth, T.~Udem, and T.~H{\"a}nsch, \enquote{Measuring the
  frequency of light with mode-locked lasers,} {\protect\JournalTitle{Optics
  Communications}} \textbf{172}, 59 -- 68 (1999).

\bibitem{Udem2002_metrology}
T.~Udem, R.~Holzwarth, and T.~W. H{\"a}nsch, \enquote{Optical frequency
  metrology,} {\protect\JournalTitle{Nature}} \textbf{416}, 233--237 (2002).

\bibitem{Deschenes_oe2010_cwdcs}
J.-D. Deschênes, P.~Giaccari, and J.~Genest, \enquote{Optical referencing
  technique with cw lasers as intermediate oscillators for continuous full
  delay range frequency comb interferometry,} {\protect\JournalTitle{Opt.
  Express}} \textbf{18}, 23358--23370 (2010).

\bibitem{Ideguchi2014}
T.~Ideguchi, A.~Poisson, G.~Guelachvili, N.~Picqu{\'e}, and T.~W. H{\"a}nsch,
  \enquote{Adaptive real-time dual-comb spectroscopy,}
  {\protect\JournalTitle{Nature Communications}} \textbf{5}, 3375 EP -- (2014).
  Article.

\bibitem{Truong_oe16_freqref}
G.-W. Truong, E.~M. Waxman, K.~C. Cossel, E.~Baumann, A.~Klose, F.~R.
  Giorgetta, W.~C. Swann, N.~R. Newbury, and I.~Coddington, \enquote{Accurate
  frequency referencing for fieldable dual-comb spectroscopy,}
  {\protect\JournalTitle{Opt. Express}} \textbf{24}, 30495--30504 (2016).

\bibitem{Zhao_oe2016_freerundcs}
X.~Zhao, G.~Hu, B.~Zhao, C.~Li, Y.~Pan, Y.~Liu, T.~Yasui, and Z.~Zheng,
  \enquote{Picometer-resolution dual-comb spectroscopy with a free-running
  fiber laser,} {\protect\JournalTitle{Opt. Express}} \textbf{24}, 21833--21845
  (2016).

\bibitem{leibrandt_rsi_2015}
D.~R. Leibrandt and J.~Heidecker, \enquote{An open source digital servo for
  atomic, molecular, and optical physics experiments,}
  {\protect\JournalTitle{Review of Scientific Instruments}} \textbf{86}, 123115
  (2015).

\bibitem{dpll_github}
J.-D. Deschênes, Digital PLL code base, see
  https://github.com/jddes/Frequency-comb-DPLL.

\bibitem{Leibrandt_opex2011_carcomb}
D.~R. Leibrandt, M.~J. Thorpe, J.~C. Bergquist, and T.~Rosenband,
  \enquote{Field-test of a robust, portable, frequency-stable laser,}
  {\protect\JournalTitle{Opt. Express}} \textbf{19}, 10278--10286 (2011).

\bibitem{sinclair_compact_comb}
L.~C. Sinclair, J.-D. Deschênes, L.~Sonderhouse, W.~C. Swann, I.~H. Khader,
  E.~Baumann, N.~R. Newbury, and I.~Coddington, \enquote{Invited article: A
  compact optically coherent fiber frequency comb,}
  {\protect\JournalTitle{Review of Scientific Instruments}} \textbf{86}, 081301
  (2015).

\bibitem{Lezius_optica2016_spacecomb}
M.~Lezius, T.~Wilken, C.~Deutsch, M.~Giunta, O.~Mandel, A.~Thaller,
  V.~Schkolnik, M.~Schiemangk, A.~Dinkelaker, A.~Kohfeldt, A.~Wicht,
  M.~Krutzik, A.~Peters, O.~Hellmig, H.~Duncker, K.~Sengstock,
  P.~Windpassinger, K.~Lampmann, T.~H\"{u}lsing, T.~W. H\"{a}nsch, and
  R.~Holzwarth, \enquote{Space-borne frequency comb metrology,}
  {\protect\JournalTitle{Optica}} \textbf{3}, 1381--1387 (2016).

\bibitem{Sinclair_14_opex_outdoorcomb}
L.~C. Sinclair, I.~Coddington, W.~C. Swann, G.~B. Rieker, A.~Hati, K.~Iwakuni,
  and N.~R. Newbury, \enquote{Operation of an optically coherent frequency comb
  outside the metrology lab,} {\protect\JournalTitle{Opt. Express}}
  \textbf{22}, 6996--7006 (2014).

\bibitem{develyntf_1994_olddigitallock}
L.~{D'Evelyn}, L.~{Hollberg}, and Z.~B. {Popovic}, \enquote{A cpw phase-locked
  loop for diode-laser stabilization,} in \emph{1994 IEEE MTT-S International
  Microwave Symposium Digest (Cat. No.94CH3389-4),}  (1994), pp. 65--68 vol.1.

\bibitem{cacciapuoti_rsi2005_phaselock}
L.~Cacciapuoti, M.~de~Angelis, M.~Fattori, G.~Lamporesi, T.~Petelski,
  M.~Prevedelli, J.~Stuhler, and G.~M. Tino, \enquote{Analog+digital phase and
  frequency detector for phase locking of diode lasers,}
  {\protect\JournalTitle{Review of Scientific Instruments}} \textbf{76}, 053111
  (2005).

\bibitem{herman_2018pra_opttiming}
D.~Herman, S.~Droste, E.~Baumann, J.~Roslund, D.~Churin, A.~Cingoz, J.-D.
  Deschênes, I.~H. Khader, W.~C. Swann, C.~Nelson, N.~R. Newbury, and
  I.~Coddington, \enquote{Femtosecond timekeeping: Slip-free clockwork for
  optical timescales,} {\protect\JournalTitle{Phys. Rev. Applied}} \textbf{9},
  044002 (2018).

\bibitem{Bluestone_oe17_cwcontrol}
A.~Bluestone, A.~Jain, N.~Volet, D.~T. Spencer, S.~B. Papp, S.~A. Diddams,
  J.~E. Bowers, and L.~Theogarajan, \enquote{Heterodyne-based hybrid controller
  for wide dynamic range optoelectronic frequency synthesis,}
  {\protect\JournalTitle{Opt. Express}} \textbf{25}, 29086--29097 (2017).

\bibitem{comm_combs}
{Known suppliers of digital laser control systems include: AOSense, Inc., IMRA
  America, Inc., Menlo Systems GmbH, TOPTICA Photonics AG, and Vescent
  Photonics, LLC. We recognize this may only be a partial list, and other
  companies or vendors may be able to provide equivalent digital control
  systems.}

\bibitem{rp_website}
{Instrumentation Technologies, LLC, Red Pitaya, see https://www.redpitaya.com}.

\bibitem{tourignyplante_rsi_2018}
A.~Tourigny-Plante, V.~Michaud-Belleau, N.~Bourbeau~Hébert, H.~Bergeron,
  J.~Genest, and J.-D. Deschênes, \enquote{An open and flexible digital
  phase-locked loop for optical metrology,} {\protect\JournalTitle{Review of
  Scientific Instruments}} \textbf{89}, 093103 (2018).

\bibitem{luda_2019_rsi}
M.~A. Luda, M.~Drechsler, C.~T. Schmiegelow, and J.~Codnia, \enquote{Compact
  embedded device for lock-in measurements and experiment active control,}
  {\protect\JournalTitle{Review of Scientific Instruments}} \textbf{90}, 023106
  (2019).

\bibitem{hansel_2017_menlo}
W.~H\"ansel, H.~Hoogland, M.~Giunta, S.~Schmid, T.~Steinmetz, R.~Doubek,
  P.~Mayer, S.~Dobner, C.~Cleff, M.~Fischer, and R.~Holzwarth, \enquote{All
  polarization-maintaining fiber laser architecture for robust femtosecond
  pulse generation,} {\protect\JournalTitle{Applied Physics B}} \textbf{123},
  41 (2017).

\bibitem{diddams_2000_link}
S.~A. Diddams, D.~J. Jones, J.~Ye, S.~T. Cundiff, J.~L. Hall, J.~K. Ranka,
  R.~S. Windeler, R.~Holzwarth, T.~Udem, and T.~W. H\"ansch, \enquote{Direct
  link between microwave and optical frequencies with a 300 thz femtosecond
  laser comb,} {\protect\JournalTitle{Phys. Rev. Lett.}} \textbf{84},
  5102--5105 (2000).

\bibitem{Diddams_2001_hgclock}
S.~A. Diddams, T.~Udem, J.~C. Bergquist, E.~A. Curtis, R.~E. Drullinger,
  L.~Hollberg, W.~M. Itano, W.~D. Lee, C.~W. Oates, K.~R. Vogel, and D.~J.
  Wineland, \enquote{An optical clock based on a single trapped 199hg+ ion,}
  {\protect\JournalTitle{Science}} \textbf{293}, 825--828 (2001).

\bibitem{washburn_hnlfoctave}
B.~R. Washburn, S.~A. Diddams, N.~R. Newbury, J.~W. Nicholson, M.~F. Yan, and
  C.~G. J{\o}rgensen, \enquote{Phase-locked, erbium-fiber-laser-based frequency
  comb in the near infrared,} {\protect\JournalTitle{Opt. Lett.}} \textbf{29},
  250--252 (2004).

\bibitem{Nishida_2003elett_nttppln}
Y.~{Nishida}, H.~{Miyazawa}, M.~{Asobe}, O.~{Tadanaga}, and H.~{Suzuki},
  \enquote{Direct-bonded qpm-ln ridge waveguide with high damage resistance at
  room temperature,} {\protect\JournalTitle{Electronics Letters}} \textbf{39},
  609--611 (2003).

\bibitem{menlo_cleo2017_ceo}
W.~H\"ansel, M.~{Giunta}, M.~{Lezius}, M.~{Fischer}, and R.~{Holzwarth},
  \enquote{Electro-optic modulator for rapid control of the carrier-envelope
  offset frequency,} in \emph{2017 Conference on Lasers and Electro-Optics
  (CLEO),}  (2017), pp. 1--2.

\end{thebibliography}

\end{document}